# Thermodynamic origins of multimode spinodal decomposition in multicomponent alloys

Pravan Omprakash

Independent Researcher, contact: pravanop@gmail.com

**Abstract.** In a binary alloy, spinodal decomposition has one composition direction in which the free energy can fall. A multicomponent alloy has more directions available, yet its elements usually separate together in a single, pseudobinary mode. Why does this happen, and what would allow two independent directions to become unstable? We examine these questions using binary interaction parameters and the composition-constrained curvature of the regular-solution free energy. Across equimolar BCC quaternaries, one unstable mode is the most common outcome at 500 K, while 15.2% have at least two. The tendency of unlike elements to mix helps explain the preference for one mode. We then show how particular arrangements of interactions, including a balanced repulsive triangle and an attractive-repulsive pair, guarantee two downhill directions. When those directions have similar strengths, simplified phase-field models produce a broad texture of locally different solid solutions rather than two chemical blocks. A phase-specific screen identifies five provisional alloys with nearly balanced modes, with Cu–In–Ir–Ti providing an example of this possibility. Thus, we establish a theoretical framework for engineering microstructures via spinodal decomposition derived from basic binary interaction tendencies.

## Introduction

Spinodal decomposition in alloys occurs when a homogeneous solid solution becomes spontaneously unstable to small composition fluctuations[1]. Through spinodal decomposition the alloy develops composition modulations throughout its volume, creating a fine-scale pattern of chemically distinct regions. These patterns can generate local environments and interfaces on length scales that are difficult to obtain by conventional processing. Subsequent nucleation can lead to microstructures that lead to enhanced strength, homogenous property distribution and potentially new functionality due to the high interfacial area and fine scale domain sizes. In binary alloys, conditions for spinodal decomposition as a function of composition and temperature can be obtained mathematically, when considering only the bulk chemical energy[2]. Subsequent phase field simulations take into consideration the strain mismatch between domains, the gradient energy across the interfaces and the mobilities of elements to separate into two distinct regions[3]. However, the bulk chemical energy provides the driving force for decomposition, with the spinodal boundary hidden inside the miscibility gap of binary alloys whose constituent elements do not like to mix together. For multicomponent alloys, the higher dimensionality makes the picture substantially more complex and interesting.

In ternary systems, as a starting point, the homogenous solid solution can reduce its energy across two different directions, rather than the constraint of one direction in binary systems. Morral and Cahn[4] showed the different type of spinodal boundaries that can occur based on the sign of binary interaction parameters between the three elements, indicating the preference of pairs of elements to mix/de-mix. Kadirvel et al.[5] developed a CALPHAD-based Hessian treatment to identify unstable composition changes in multi-principal-element alloys, while Raymundo et al. extended the framework to include coherent strain. These frameworks allow for the mathematical analysis of spinodal decomposition when provided with a suitable free energy function. Using the regular solution model derived from liquid binary interaction parameters Luan et al.[6] argued that decomposition in multicomponent alloys behave in a pseudo-binary or single-mode decomposition; essentially many elements collate to form chemical blocks that move in two different directions, i.e., an alloy A-B-C-D spinodally decomposes to form AB|CD domains, rather than a A|BC|D domains. Their work thus leads to three simple questions:

1) Why do the available energy-lowering directions in multicomponent energy space most often collapse into one unique direction?
2) What should the arrangement of chemical interactions be to allow for the seemingly special multi-direction instability?
3) When the so-called multimode instability is possible, what does the resulting composition texture look like?

Here, we address those questions using the large binary-interaction dataset assembled by Chen et al.[7], and the framework established by Kadirvel et al.[5] to calculate composition-constrained free-energy curvatures across multicomponent alloys and study decomposition directions. We connect the prevalence of a dominant mode to the distribution of interaction parameters containing a negative mean. We then derive the special conditions, in the form of weighted interaction motifs, that guarantee at least two unstable directions in regular-solution alloys containing four or more elements. These constructions make the origin of multimode instability visible in the pair interactions themselves, without having to calculate a computed eigenvector. These motifs also guarantee multimode decomposition in any multicomponent system with more than four elements. Finally, a phase-specific screening of 91,000 quaternary alloys yield only five provisional candidates with nearly balanced two unstable directions. Simple phase-field calculations illustrate how multimode decomposition leads to non-collinear chemical segregation, resulting in what could be described as a continuous distribution of compositionally different solid solutions. Thus, we establish a theoretical framework for engineering microstructures via spinodal decomposition derived from basic binary interaction tendencies.

**Mathematical foundation for chemical spinodal decomposition**. We start with defining the Gibbs free energy,$G(x,T)$, of a multiple principal element alloy (MPEA)—assuming a regular solution model—as:

$$G(x,T) = \Sigma_{i<j}\Omega_{ij}x_ix_j + k_BT\Sigma_i x_i ln x_i + \Sigma_i x_i g_i^o(T), \qquad \text{Equation (1)}$$

where, $x$ is the mole fraction of each element $i$, $T$ is the temperature in K, $k_B$ is the Boltzmann constant, $g_i^o(T)$ is the absolute energy of the pure element as a function of temperature, and $\Omega_{ij}$ is the interaction parameter obtained from Chen et al.'s DFT dataset. Positive $\Omega_{ij}$ penalizes mixing of the corresponding pair; negative values favor unlike pairs and therefore ordering, and values close to 0 imply random mixing. These interaction parameters are lattice specific, and the article uses BCC parameters as the model to establish its conclusions. However, for the final screening of quaternary candidates the appropriate lower energy lattice between BCC, HCP and FCC is used.

Next, we determine the Hessian, $\boldsymbol{H}$, which determines the curvature of the free energy surface. We first define the interaction parameter matrix, $\boldsymbol{\Omega_{m \times m}}$ , where m is the number of elements minus 1. Because the sum of mole fractions have to sum up to one, the molar fraction of one element is constrained and thus not a free variable. Let us then define an orthonormal basis for composition changes, $Q$, satisfying $1^TQ = 0$. We can then express any change in composition, $\delta x = Q.z$, where $z$ is the vector representing independent changes in composition, that conserves the sum of fractions. From Equation 1, we can derive the $\boldsymbol{H}$—taking into the constraints of molar fractions—to be:

$$\boldsymbol{H_T} = Q^T\left[\boldsymbol{\Omega} + \mathrm{k_B}T\mathrm{diag}\left(\frac{1}{\mathrm{x_i}}\right)\right]Q. \qquad \text{Equation (2)}$$

We can then define the change in energy along $z$ as:

$$\delta g_2 = \frac{1}{2}z^T\boldsymbol{H_T}z. \qquad \text{Equation (3)}$$

Diagonalization of $\boldsymbol{H_T}$, allows us to express the local landscape as:

$$\delta g_2 = \frac{1}{2}\Sigma_a \lambda_a \eta_a^2, \qquad \text{Equation (4)}$$

where, $\lambda_a$ are the eigenvalues of the hessian, and the $\eta$ are the corresponding eigenvectors. A 3-component alloy has 2 independent mole fractions, resulting in 2 corresponding $\lambda$, the eigenmodes (referred henceforth simply as modes), as shown in Figure 1(a)-(b). The sign of the eigenvalues specifies whether the energy change is in a downhill (energy-lowering) or uphill (energy-increasing) direction. Figure 1(a)-(b), contrasts a saddle with one negative curvature and a landscape with two. In the latter, every nonzero combination within the unstable plane lowers the quadratic energy. Thus, multimode decomposition is a two-dimensional chemical freedom, not simply a choice between two proposed binary partitions.

**Frequency of multimode decomposition.** Using this framework, we enumerate all $\binom{40}{4} = 91{,}390$ quaternaries from the supplied BCC binary-interaction database. In the supplementary we exhibit that FCC and HCP interactions also follow a similar pattern. At 0 K, 4.1% have no unstable modes, 65.9% have one and 30% have at least two, as shown in Figure 1(c). At 500 K, these fractions become 8.9%, 75.9% and 15.2% respectively. Thus, aligning with Luan et al.[6], one-mode instability is therefore the predominant local decomposition geometry in this substantive alloy set, while multimode instability occupies a substantial minority.

At equimolar quaternary composition, ideal entropy adds the same positive curvature to every mode, hence we can decompose the $\boldsymbol{H}$ to $\boldsymbol{H_T} = \boldsymbol{H_O} + \tau\boldsymbol{I}, \tau = 4k_BT$, where $\boldsymbol{H_O}$ is the part of the hessian describing the 0K contribution to the mode analysis.

Heating lifts the weaker negative curvature to zero first, resulting in the 30% two-mode to reduce to 15.2% at 500 K. This leads to the conclusion that even within two-mode unstable compounds, there is a disparity between the two surviving curvatures: a fact that was analyzed and noted by Luan et al.[6] too. To measure the relative strength, we define $R = \frac{|\lambda_1|}{|\lambda_2|}$, with an ordered mode set, $\lambda_1 \leq \lambda_2 < 0$. Among the 13,519/91,390 systems with exactly two unstable modes at *both* temperatures, the median *R* decreases from 0.199 at 0 K to 0.099 at 500 K, as shown in Figure 1(d). The vast majority of the two-mode population thus still has a strongly preferred downhill direction. Thus, between the two differently realized theoretical reports, this work and Luan et al., it is safe to say that the pseudo-binary decomposition of is a property of MPEAs.

A bold claim on the natural description of alloys could be made at this juncture: nature seems to choose an unambiguous direction through which it lowers energy, rather than having multiple possible directions to lower its energy. Accounting for strain, diffusion and interface energetics could change this picture in either way: the preference for single-mode instability could increase further, as argued by Luan et al.[6], or two-mode instabilities could become more preferred. A rigorous theoretical description of strain energy preference to either case needs to be established before such a conclusion can be drawn. Such an analysis is beyond the scope of this work but is recognized as a vital route for future work.

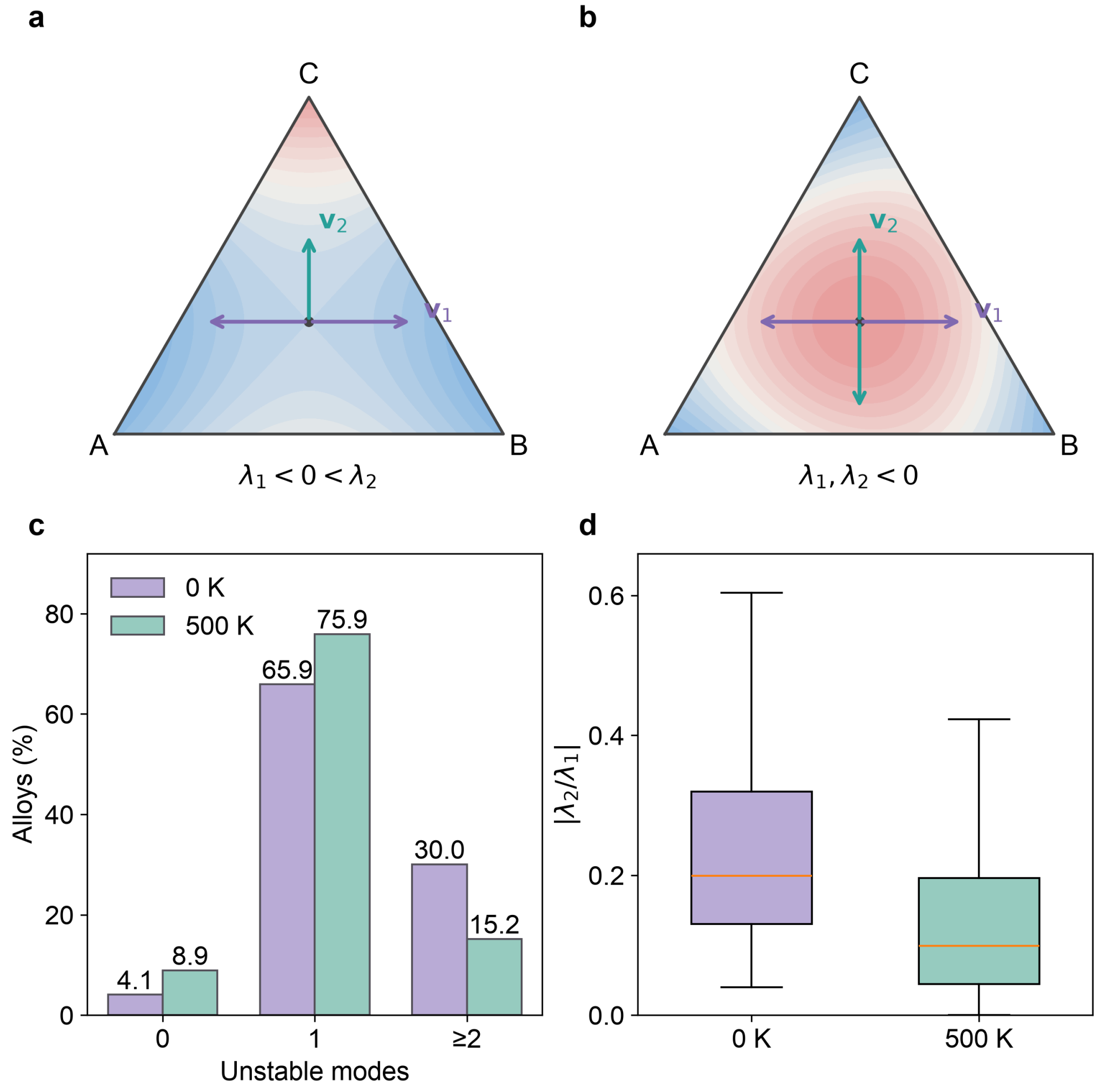

**Figure 1. The dimensionality of spinodal instability.** (a,b) Schematic ternary composition landscapes with one and two negative curvatures; arrows mark the local eigenvector directions. (c) Unstable-mode counts for 91,390 equimolar BCC quaternaries at 0 and 500 K. (d) Curvature ratio R = |λ2|/|λ1| for the same 13,519 systems with exactly two negative modes at both temperatures. Boxes show quartiles and medians; whiskers extend to observations within 1.5 interquartile ranges. Ratios describe curvature strength rather than evolved composition amplitude.

## Why do the available energy-lowering directions in multicomponent energy space most often collapse into one unique direction?

Having established the basis for our questions, we proceed to answer the first question. There are almost double the number of single-mode instabilities compared to two-mode instabilities. Since the $\boldsymbol{H}$ depends on the interaction parameters as derived in Equation 2, we ask if the interaction parameters of each alloy have a property in their distribution that leads to an abundance in single-mode instabilities. Indeed, the interaction distribution of the entire BCC set provides a first explanation for this preference. The 780 BCC pair parameters have a negative mean of -0.415 eV and a standard deviation of 1.219 eV. The real distribution is shown as the purple histogram in Figure 2(a).

We compare this distribution with ten synthetic 40-element tables drawn from a Gaussian with the same mean and spread, and ten drawn from a zero-mean Gaussian of the same spread. Each synthetic pair is drawn once and reused in every quaternary containing it. This preserves the combinatorial sampling structure of the real distribution. The green curve in Figure 2(a) corresponds to the fitted Gaussian, and the zero-mean Gaussian is shown in Figure 2(b). The fitted Gaussian approximately reproduces the one-mode and multimode fractions, whereas the zero-mean ensemble places half of the quaternaries in each of the two categories—expected from a normal distribution. This procedure of fitting Gaussians with means and standard deviations, is reproduced for a large grid of means and spreads, ranging from large negative to large positive values (for the mean), as shown in Supplementary Section 1.

The negative mean found organically in the interaction parameter terms therefore helps explain the predominance of single-mode instability. Simply put, more elements like to mix together, than segregate. Thus, in majority of the quaternaries, two groups can be formed, with *at least* one element each, and both these groups can separate along one unstable direction. The elements within these groups are sufficiently *happy* with each other, and therefore do not need another unique unstable direction, i.e., don't require multimode decomposition. This statistical picture naturally motivates the next question.

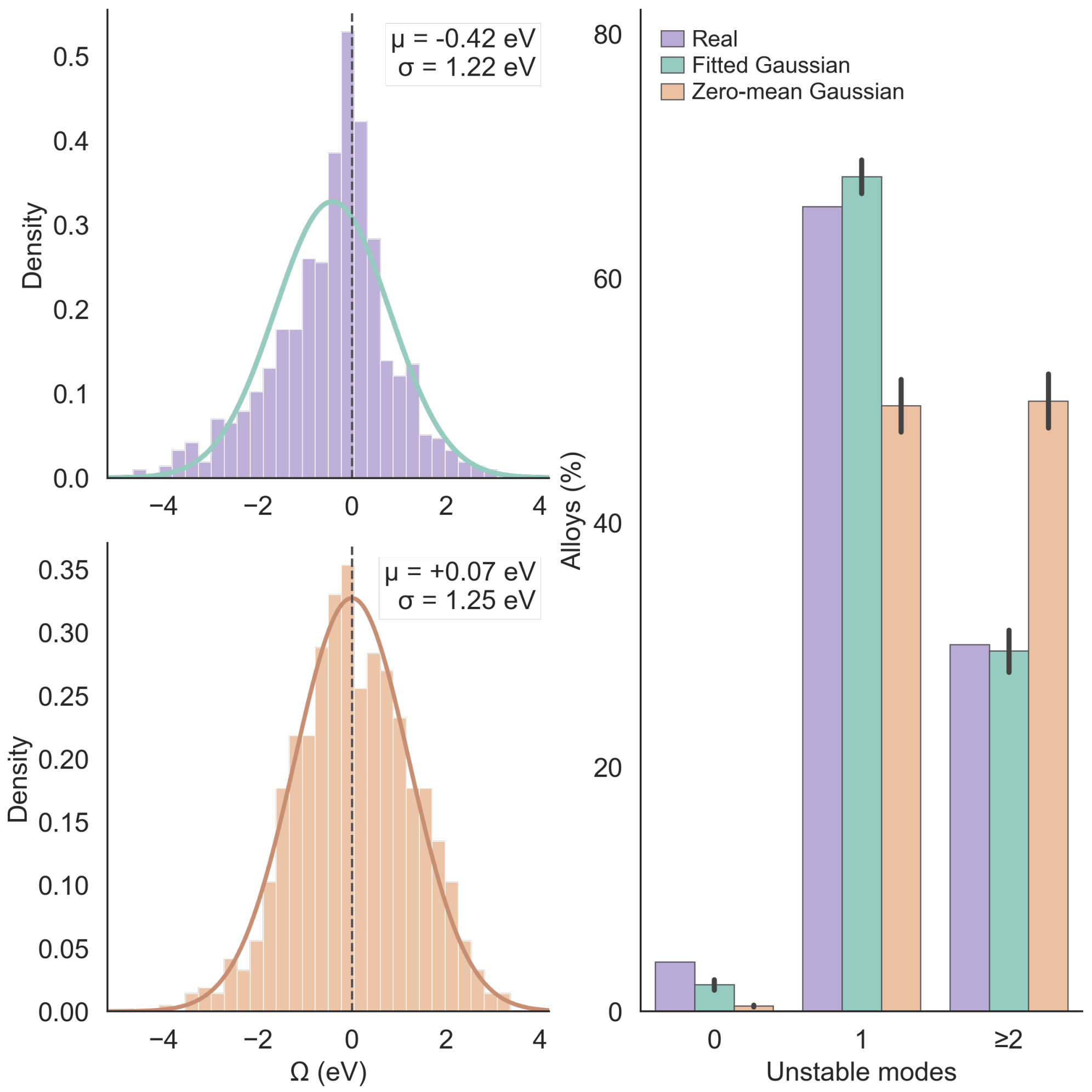

**Figure 2 Interaction statistics favor a dominant direction.** The real BCC pair distribution and its Gaussian fit are compared with a zero-mean Gaussian of the same standard deviation. The bar plot gives 0 K mode fractions for real, fitted-Gaussian, and zero-mean-Gaussian interaction tables. Synthetic bars average ten complete 40-element tables; error bars show their standard deviation.

## What should the arrangement of chemical interactions be to allow for the seemingly special multi-direction instability?

We can assess the interaction matrix, $\mathbf{\Omega}$, with a definite strategy to find *motifs* in the interactions to guarantee multimode decomposition. The exact condition can also be derived as presented in Supplementary Section 2 but is intuitively opaque. We, therefore, present two motifs that are easy to understand and have guaranteed conditions based on interaction magnitude and sign that guarantee multimode decomposition. These two motifs account for ~85% of the 13,855 quaternaries two-mode instabilities at 500 K. Owing to the fact that these motifs form a subset of any interaction matrix in quinaries, senaries etc., they form a substantive—and intuitive—basis for multimode decomposition.

The common principle of arriving at these motifs and the conditions that they must satisfy is generic across all four motifs. We choose two independent, conserved composition changes, $p$ and $q$, using the interactions rather than Hessian eigenvectors. These two orthogonal vectors are $p$ and $q$, define a 2-D plane in the energy surface, where every change in their plane is $\delta x = ap + bq$. We can then define a reduced Hessian, $\boldsymbol{H}_{\boldsymbol{T}}^{\boldsymbol{reduced}}$, describing the curvature within that plane, derived from Equation (2):

$$\boldsymbol{H}_{\boldsymbol{T}}^{\boldsymbol{reduced}} = \begin{bmatrix} p^T \Omega p + \tau & p^T \Omega q \\ \mathrm{p}^{\mathrm{T}} \Omega q & q^T \Omega \mathrm{q} + \tau \end{bmatrix} \qquad \text{Equation (5)}$$

Within this plane, if two directions are downhill, then we have multimode decomposition. The third direction could also be uphill or downhill; hence this method provides sufficient but not necessary conditions. If we require two directions to be downhill, or have two negative eigenvalues, from $\boldsymbol{H}_{\boldsymbol{T}}^{\boldsymbol{reduced}}$, then two simple conditions need to be satisfied:

$$p^T \Omega p + \tau < 0; (p^T \Omega p + \tau)(p^T \Omega p + \tau) > \ (p^T \Omega q)(p^T \Omega q) \qquad \text{Equation (6)}$$

The second condition is essentially stating that the determinant of the matrix has to be positive. If we are able to find this plane within our multi-dimensional free energy surface, then two-mode decomposition is guaranteed. The plane criterion is a sufficient, two-dimensional counterpart to the exact three-dimensional test for multimode instability, as derived in Supplementary Sec. II. We now use it to construct representative interaction motifs. These motifs are not unique descriptions of an alloy: one interaction set may support several downhill planes, with different curvatures and preferred directions. For the motif counts, we assign each alloy to the first criterion it satisfies in the stated order, solely to avoid double counting. The resulting rows measure the additional alloys certified by each motif, not the strength of that motif or a unique decomposition pathway.

1. **<u>Balanced Repulsive Triangle</u>**

Let us consider an interaction graph of *ABCD* with three mutually repulsive interactions, $\Omega_{AB}, \Omega_{BC}, \Omega_{AC} > 0$, as shown in Figure 3(a). The fourth component, *D*, can have any interaction, and its composition can be considered to be fixed, $\delta x_d = 0$. The allowed changes then form a 2D plane, $\delta x_A + \delta x_B + \delta x_C = 0$. We can now choose $p$ and $q$ such that they form convenient basis vectors that span this whole plane. Since A and B don't want to mix, they will separate, so we can choose $p = \frac{[1,-1,0,0]}{\sqrt{2}}$, such that it describes an enrichment in A and subsequent depletion in B. An orthonormal $q = \frac{[1,1,-2,0]}{\sqrt{6}}$ then gives us the basis vectors. The plane criterion from Equation (6) is now in play:

$$\Omega_{AB} > \tau \qquad \text{Equation (7)}$$

We define the interaction parameters to be positive, but one interaction has to be strong even at the higher temperatures for two downhill directions to be possible.

For the second condition, we define a convenient quantity, $\Delta$:

$$\Delta = \frac{2(\Omega_{AB}.\Omega_{BC} + \Omega_{BC}.\Omega_{AC} + \Omega_{AB}.\Omega_{AC}) - (\Omega_{AB}^2 + \Omega_{BC}^2 + \Omega_{AC}^2)}{3} \qquad \text{Equation (8)}$$

$$\Delta - 2\tau(\Omega_{\mathrm{AB}} + \Omega_{BC} + \Omega_{AC}) + 3\tau^2 > 0 \qquad \text{Equation (9)}$$

At 0K, $\Delta > 0$, which essentially describes that the triangle should be sufficiently balanced, if either of three interaction parameters is overwhelmingly positive, then only separation direction will become unstable. The interactions with *D* do not enter this sufficient construction, they could create a third unstable mode but cannot influence these two directions due to the plane criterion.

In essence this condition represents a *frustrated* triangle in interaction space; a useful analogue can be drawn to the occurrence of non-collinear magnetism in triangular lattices, where all nearest-neighbor atoms prefer to have anti-parallel alignment with each other creating magnetic frustration. The triangle motif and the condition reproduces two-mode instability for 7642 (55.2%) of the 13,855 alloys at 500 K and provides the clearest example where competing chemical preferences support a downhill plane rather one separation axis.

## 2. Attractive and Repulsive pair

Now, let us consider the case where, $\Omega_{AB} > 0, \Omega_{CD} < 0$, as shown in Figure 3(b). Then, $p = \frac{[1,-1,0,0]}{\sqrt{2}}$ and $q = \frac{[1,1,-1,-1]}{2}$, where the second direction implies that *AB* moves against *CD*, on account of the fact that CD prefer to move together. We can define sum and imbalance between the four interactions between the groups as:

$$\Sigma = \Omega_{AC} + \Omega_{AD} + \Omega_{BC} + \Omega_{BD}, \Gamma = (\Omega_{BC} + \Omega_{\mathrm{BD}}) - (\Omega_{AC} + \Omega_{AD}), \qquad \text{Equation (10)}$$

yielding:

$$4(\Omega_{AB} - \tau)(\Omega_{AB} + \Omega_{CD} + 2\tau - \Sigma) < \Gamma^2 \qquad \text{Equation (11)}$$

The interpretation is intuitive: A-B can exchange with a sufficiently strong interaction, while the strong *CD* attraction can drive the groups to segregate, provided that the imbalance within their interactions is not too strong. In other words, the *CD* attraction has to be sufficiently strong to create two unstable directions. This motif supplies two downhill paths for an additional 4090 alloys of the 6,213 alloys that do not contain the repulsive triangle motif.

Three other motifs are also discussed in Supplementary Section II. Interestingly, 72 alloys at 0K have **all** negative interactions but still have two-mode instabilities; they do not survive higher temperatures though. Together, the five motifs account for about 93% of the total two-mode alloys, providing a substantive basis for understanding the interaction patterns that lead to two-mode instabilities. The exact condition, described in Supplementary Section II can always be used to determine the necessary condition for any system whose interaction parameters are known.

Now that we have understood the criterions for having multimode decomposition, let us understand how the microstructures differ due to the different number and strength of the unstable modes.

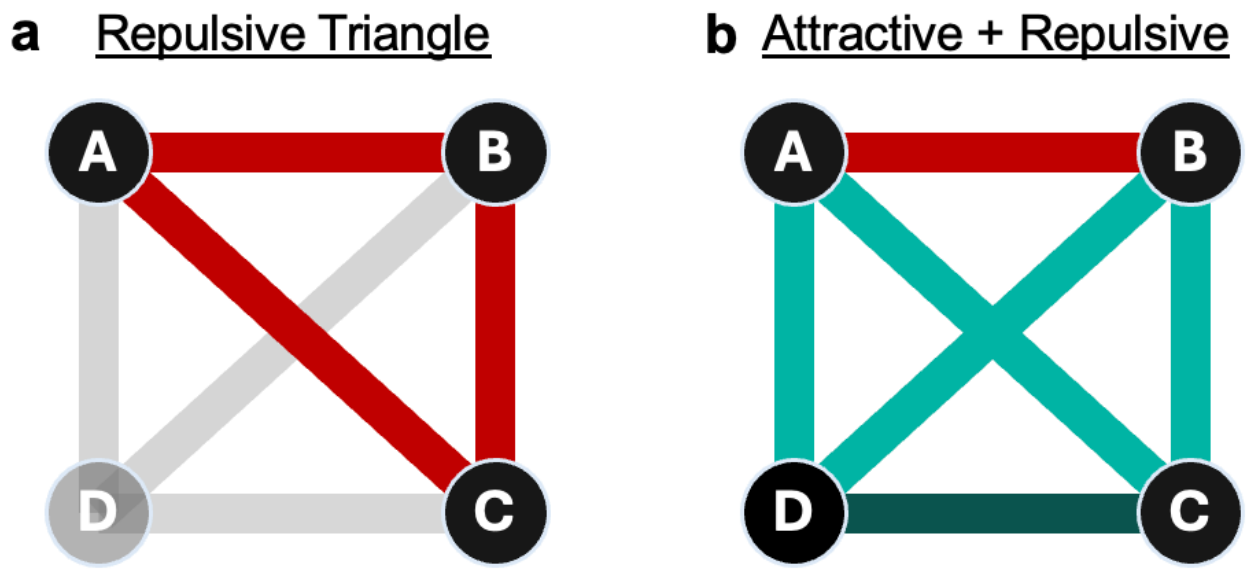

**Figure 3. Two sufficient interaction motifs.** (a) A balanced repulsive triangle supplies two conserved downhill changes within $ABC$, regardless of the three interactions with $D$. (b) A repulsive $AB$ exchange can coexist with separation of the cohesive $CD$ pair from $AB$. Warm edges indicate repulsion, cool edges attraction, and gray edges are unrestricted in the triangle motif. Edge signs illustrate the constructions; the weighted inequalities in Eqs. (8) and (11) provide the guarantees.

## When the multimode instability is possible, what does the resulting composition texture look like?

So far, we have disregarded the relative strength of the two unstable modes. Now, let us take systems with two unstable modes, and look at their quadratic energy change, derived from Equation (4):

$$\delta g_2 = \frac{1}{2}\lambda_a \eta_a^2 + \frac{1}{2}\lambda_b \eta_b^2;\ \lambda_a, \lambda_b < 0. \qquad \text{Equation (12)}$$

Figure 4 illustrates the distinction when the relative strengths are different using conserved phase-field toy models. All runs use the same compositional noise, scalar mobility, isotropic gradient penalty, quartic stabilization and elapsed time. The details of the phase field simulations are detailed in Supplementary Section IV. The interaction matrices are constructed to illustrate the two contrasting groups of elements, AC|BD. With a single-mode instability, as shown in Fig. 4(a), the changes satisfy $x_A - x_B = x_C - x_D$, as shown in the rightmost panel. This simulation hence shows a pseudo-binary decomposition, wherein two solid solutions AC and BD form an interface.

In the second case, both unstable modes are nearly equal in magnitude, at a fixed amplitude $\rho$, we can write $(\eta_1, \eta_2) = \rho(cos\theta, sin\theta)$, gives:

$$\delta g_2 = \frac{\rho^2}{4}[(\lambda_a + \lambda_b) + (\lambda_b - \lambda_a)cos2\theta] \qquad \text{Equation (13)}$$

Thus, the curvature difference is directly the quadratic angular preference between the vectors. The point cloud in the rightmost panel of Figure 4(b) shows that the A and B separate in a different direction than C and D, creating a near spherical cloud described by Equation (13). This leads to a phase field that is a continuous distribution of solid solutions with *A-rich* regions, *B-rich* regions, *C-rich* regions and *D-rich* regions. A fair question to ask at this juncture is how this microstructure different from a purely segregated elemental mixture; the key difference is the large interfacial area that can form on account of spinodal decompositions. Some of these regions could be more prone to intermetallic precipitation than others and thus could give rise to potentially interesting properties. Thus, the resulting texture provide a conceptual route to controlling the coexistence and spatial arrangement of several local compositions with one decomposing material.

The third case exhibits what happens when the second unstable mode is not as strong as the first one, thus leading to the collapse of the multimode decomposition. *C* and *D* do not decompose separately, leading to two domains which is effectively *ACD* and *BCD* rich respectively. Note that this is also a useful microstructure, and thus even if two modes are present with extremely low strength ratio, heat treatment at different temperatures can lead to compositionally different microstructures that can allow for control of properties.

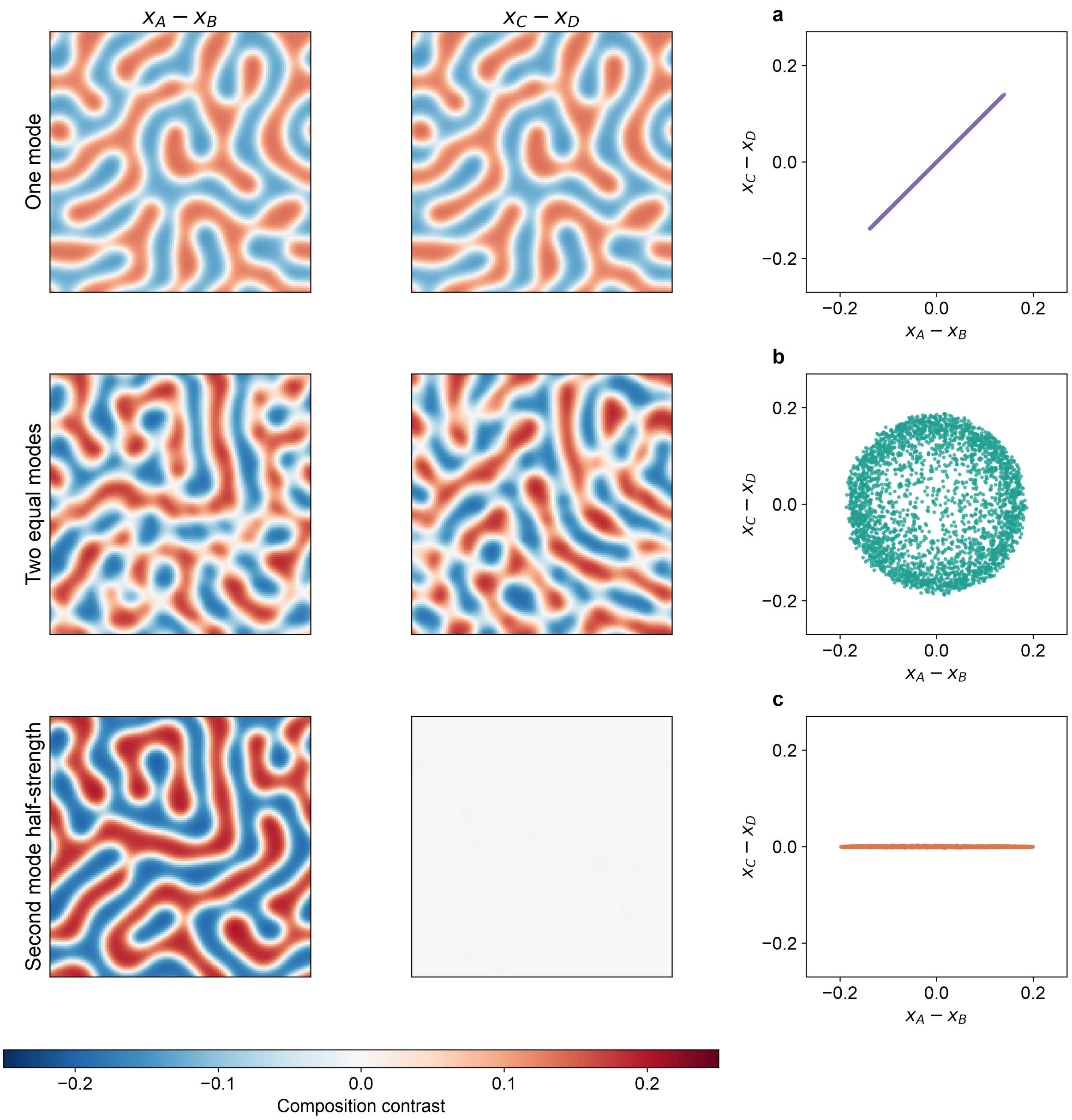

**Figure 4. Composition textures from conserved phase-field toy models.** Rows show one unstable mode, two equally unstable modes, and a second mode with half the leading negative curvature. The first two columns map $x_A - x_B$ and $x_C - x_D$ on the same scale; the third plots these contrasts for 3,000 spatial pixels. All rows use identical noise and model coefficients and are shown at reduced time 260. The half-strength second mode has not developed appreciable amplitude at this time.

We next screen equimolar quaternaries at 500 K using their lowest homogeneous regular-solution free energy among BCC, FCC, and HCP. The comparison includes the supplied unary lattice free energies and phase-specific interactions. Of 82,251 phase-evaluable systems, 9,138 have at least two unstable modes. Five have exactly two with symmetric relative curvature splitting, shown in Table 1. The best quaternary system, Cu-In-Ir-Ti is selected as an illustrative example. Two near equal unstable modes are present: 1) A Cu-In exchange, and 2) Cu-In collectively away from Ir-Ti. This candidate thus shows the repulsive-attractive pair motif discussed in the previous section (Figure 3(b)). The six BCC interactions are shown in Table 2. The modal strengths (-0.5046 vs -0.4820) are balanced despite the very different raw interaction strengths. Figure 5(a) shows the local two-mode landscape with the positive, stable third mode relaxed. We plot the tangent-subtracted free energy in Equation (14):

$$\tilde{g}(\eta_1, \eta_2) = \min_{\eta_3} \left\{ g\left( \mathbf{x}_0 + \sum_{\alpha=1}^{3} \eta_\alpha \mathbf{v}_\alpha, T \right) - g(\mathbf{x}_0, T) - \nabla g(\mathbf{x}_0, T) \cdot \sum_{\alpha=1}^{3} \eta_\alpha \mathbf{v}_\alpha \right\}. \qquad \text{Equation (14)}$$

| Alloy | Phase | $\lambda_1$ (eV) | $\lambda_2$ (eV) | D |
|---|---|---|---|---|
| **Cu–In–Ir–Ti** | **BCC** | **-0.5046** | **-0.4820** | **0.0457** |
| Cu–In–Pt–Si | HCP | -0.2883 | -0.2705 | 0.0636 |
| Cu–Rh–Ti–Y | BCC | -0.4916 | -0.4540 | 0.0796 |
| Cu–Hf–Ni–V | HCP | -0.3286 | -0.3075 | 0.0662 |
| Cr–Re–Ta–Zn | BCC | -0.4515 | -0.4153 | 0.0835 |

**Table 1. Phase-selected quaternaries with $D < 0.10$ at 500 K**. The margin is the selected homogeneous lattice free energy below the next-lowest available BCC, FCC, or HCP lattice model. The rows are ordered as in the candidate screen, not by $D$.

| Pair | Cu–In | Cu–Ir | Cu–Ti | In–Ir | In–Ti | Ir–Ti |
|---|---|---|---|---|---|---|
| **$\Omega_{ij}$ (eV)** | 0.6710 | -0.6435 | -0.2001 | -0.4733 | -0.3486 | -3.6264 |

**Table 2. BCC binary interaction parameters for Cu–In–Ir–Ti in the selected regular-solution model.** Positive values penalize mixing; negative values favor unlike pairs.

This gives rise to a Mexican hat energy landscape, commonly used to describe Goldstone bosons; however, in this context, they are more precisely termed as pseudo-goldstone behaviour in the composition plane.

We push forward this analogue from the previous comparison with triangular lattice, frustration driven magnetism. And it remains only that, a useful comparison. Composition planes with their mass conservation criterions are in no way related to bosons. The full phase-field calculation develops an extended composition cloud, as seen in Figure 5(b) and distinct elemental mapping, as seen in Figure 5(c). Notice the interlocking Cu-rich, In-rich, IrTi-rich regions, essentially floating in the equimolar solid solution matrix. These are the kind of microstructures that  multimode decomposition can enable in MPEAs.

As a possible functional direction, the spatial coexistence of Cu-rich and Ir-containing environments could let one vary neighboring catalytic sites and their interfaces while holding the average alloy composition fixed[8]. Cu and Ir are known for their optimal catalytic activity. The local surface chemistry, accessibility, and stability of those sites would have to be established separately. The present calculation only identifies a thermodynamic route to such a distribution of chemical environments, rather than a demonstrated catalytic function.

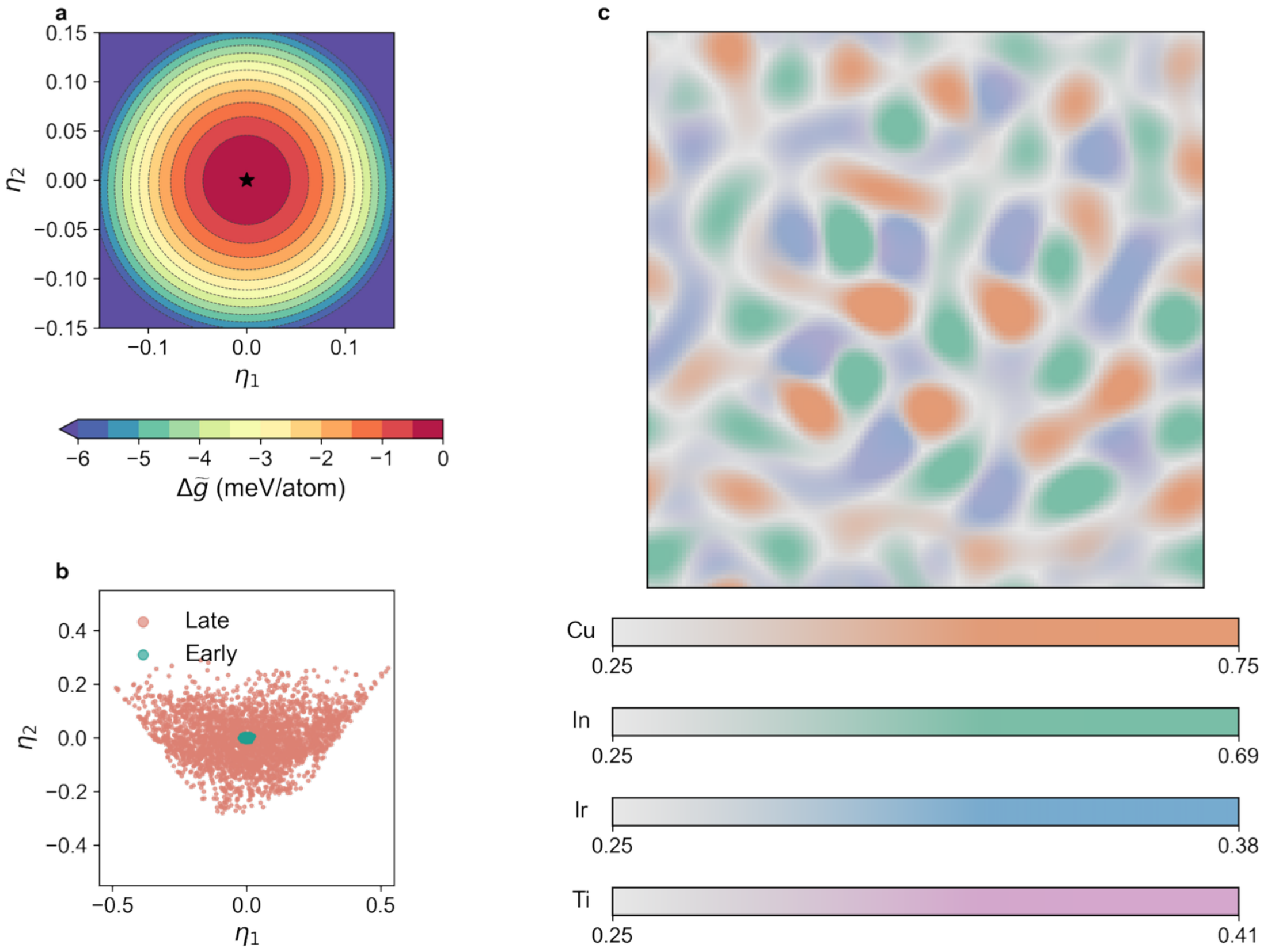

**Figure 5. A nearly balanced alloy instability: Cu–In–Ir–Ti at 500 K.** (a) Stable-mode-relaxed, tangent-subtracted regular-solution energy in the two unstable mode coordinates. (b) Composition-space scatter at reduced times 290 (early) and 500 (late). (c) Late Cu, In, Ir, and Ti fraction maps with four color scale. The selected BCC model has λ1 = −0.5046 and λ2 = −0.4820 eV.

## Conclusion

Multimode spinodal decomposition begins when the binary interactions support two independent downhill directions in composition space. In the alloys studied here, a single direction is more common, but balanced repulsive triangles and pair-exchange motifs provide simple conditions that guarantee a second mode. When the two modes have similar strengths, decomposition can generate a broad distribution of local solid-solution compositions rather than two chemical blocks. Cu–In–Ir–Ti illustrates how this freedom may appear in a multicomponent alloy. Such textures could bring Cu-rich and Ir-containing regions, and their interfaces, into close proximity, offering a possible route to varied catalytic sites within one material. The interaction motifs therefore provide a way to think about both the origin of multimode instability and the chemical landscapes it might create. Thus, we establish a theoretical framework for engineering microstructures via spinodal decomposition derived from basic binary interaction tendencies.

# Supplementary Material

## SI. Data, phase statistics, and Gaussian comparisons

The interaction dataset contains 40 elements and one value for each of their $\binom{40}{2} = 780$ pairs in each BCC, FCC, and HCP lattice model. For each lattice, all $\binom{40}{4} = 91{,}390$ equimolar quaternaries are enumerated. We form the symmetric $4 \times 4$ matrix $\mathbf{\Omega}$, set its diagonal to zero, project it onto an orthonormal basis of the three-dimensional space $\mathbf{1}^{\mathrm{T}}\delta\mathbf{x} = 0$, and count eigenvalues below $-10^{-8}$ eV. At $x_i = 1/4$, the ideal configurational term adds $\tau I_3$ with $\tau = 4k_{\mathrm{B}}T$. The lattice-specific means, spreads, and mode counts are shown in Table S1.

| Lattice | μ (eV) | σ (eV) | T (K) | 0 modes (%) | 1 mode (%) | ≥2 modes (%) |
|---|---|---|---|---|---|---|
| **BCC** | -0.415 | 1.219 | 0 | 4.06 | 65.89 | 30.04 |
| **BCC** | -0.415 | 1.219 | 500 | 8.90 | 75.94 | 15.16 |
| **FCC** | -0.260 | 1.055 | 0 | 2.45 | 65.68 | 31.87 |
| **FCC** | -0.260 | 1.055 | 500 | 7.95 | 78.96 | 13.09 |
| **HCP** | -0.303 | 1.099 | 0 | 3.16 | 62.45 | 34.39 |
| **HCP** | -0.303 | 1.099 | 500 | 7.79 | 76.85 | 15.37 |

**Table S1. Interaction statistics and equimolar quaternary mode fractions by lattice.** The last category includes both two and three negative curvatures. Phase-specific tables are analyzed separately here; Table 1 instead chooses one lowest-energy homogeneous lattice for each phase-evaluable alloy.

For Fig. 2, ten independent 40-element synthetic tables are generated for each Gaussian ensemble. A pair is sampled once within a table and reused across all quaternaries containing it. The fitted ensemble uses the BCC mean and standard deviation, while the centered ensemble sets the mean to zero and retains the BCC spread. Bars show the mean of the ten resulting mode fractions and error bars show their standard deviation. The displayed zero-mean histogram is one finite realization and consequently has a small nonzero sample mean.

To examine a broader range of distributions, we also sample six independent Gaussian interactions per quaternary over means from $-1.5$ to $1.5$ eV and standard deviations from $0.05$ to $2.0$ eV. Each grid point pools ten runs of 1,000 six-edge draws at 0 K. This grid probes the role of mean and spread without the shared-pair correlations present in a complete 40-element table. The maps in Figure S1 mark the BCC,

FCC, and HCP mean–spread coordinates. They show that shifting a distribution toward negative mean favors stability or one-mode curvature, while sufficiently broad or positive interaction distributions favor additional negative directions.

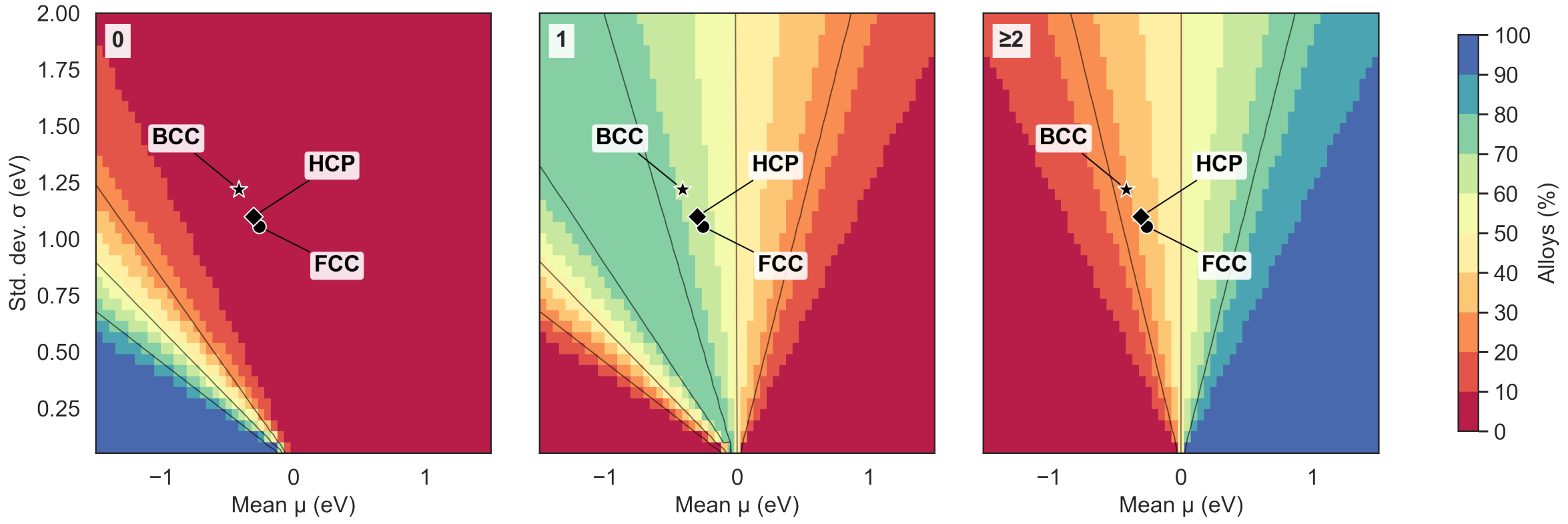

**Figure S1. Gaussian mean–spread map of 0 K quaternary mode fractions.** The three panels give percentages with zero, one, and at least two negative curvatures, respectively. Each grid point comprises 10,000 independent six-edge draws. Markers locate the empirical BCC, FCC, and HCP pair-distribution means and standard deviations; colors use discrete ten-percentage-point intervals.

## SII. Exact mode count from the interaction matrix

For a quaternary, let $K$ denote its $3 \times 3$ constrained Hessian in any complete composition coordinate basis,

$$K = \begin{pmatrix} a & b & c \\ b & d & e \\ c & e & f \end{pmatrix}. \tag{S1}$$

The basis need not be orthonormal for the *negative count* to be meaningful, although its numerical eigenvalues then depend on the coordinate scale. For $a \neq 0$, completing the first square in $\mathbf{z}^{\mathsf{T}} K \mathbf{z}$ gives

$$\mathbf{z}^{\mathsf{T}} K \mathbf{z} = a\left(z_1 + \frac{bz_2 + cz_3}{a}\right)^2 + sz_2^2 + 2tz_2z_3 + rz_3^2, \tag{S2}$$

where $s = d - b^2/a$, $t = e - bc/a$, and $r = f - c^2/a$. If $s \neq 0$, completing the remaining cross term gives

$$\mathbf{z}^{\mathsf{T}} K \mathbf{z} = a\left(z_1 + \frac{bz_2 + cz_3}{a}\right)^2 + s\left(z_2 + \frac{t}{s}z_3\right)^2 + \left(r - \frac{t^2}{s}\right)z_3^2. \tag{S3}$$

The transformed coordinates are independent because the change of variables is invertible. Sylvester's law of inertia therefore states that the number of negative eigenvalues equals the number of negative coefficients in Eq. (S3). With $\Delta_0 = 1$, $\Delta_1 = a$, $\Delta_2 = ad - b^2$, and $\Delta_3 = \det K$, these coefficients are $d_1 = \Delta_1$, $d_2 = \Delta_2/\Delta_1$, and $d_3 = \Delta_3/\Delta_2$. When no pivot vanishes, $n_-(K) \geq 2$ exactly when at least two of $d_1, d_2, d_3$ are negative.

One can make the dependence on the six pair interactions explicit without diagonalization. Eliminating $x_D = 1 - x_A - x_B - x_C$ at 0 K gives the symmetric coordinate Hessian

$$M_0 = \begin{pmatrix} -2\Omega_{AD} & \Omega_{AB} - \Omega_{AD} - \Omega_{BD} & \Omega_{AC} - \Omega_{AD} - \Omega_{CD} \\ \Omega_{AB} - \Omega_{AD} - \Omega_{BD} & -2\Omega_{BD} & \Omega_{BC} - \Omega_{BD} - \Omega_{CD} \\ \Omega_{AC} - \Omega_{AD} - \Omega_{CD} & \Omega_{BC} - \Omega_{BD} - \Omega_{CD} & -2\Omega_{CD} \end{pmatrix}. \tag{S4}$$

For example, its first square-completion coefficient is $-2\Omega_{AD}$: an $A \leftrightarrow D$ exchange is downhill when $\Omega_{AD} > 0$. The remaining coefficients determine whether other independent downhill directions exist. The eigenvalues of $M_0$ should not be compared numerically with eigenvalues of an orthonormally projected Hessian; the inertia is the same.

If a pivot vanishes, define $T_1 = \mathrm{tr}K$, $T_2 = [(\mathrm{tr}K)^2 - \mathrm{tr}(K^2)]/2$, and $T_3 = \det K$. These are the first three elementary symmetric functions of the eigenvalues. A pivot-free exact test is

$$n_-(K) \geq 2 \quad \Leftrightarrow \quad \begin{cases} T_1 < 0 \text{ or } T_2 < 0, & T_3 > 0, \\ T_1 < 0 \text{ and } T_2 > 0, & T_3 \leq 0. \end{cases} \tag{S5}$$

The test counts strictly negative curvatures, including cases with one exactly zero curvature. At finite temperature, evaluate it on the complete $K_T$ from Eq. (2). The two-dimensional motif tests in the main text guarantee $n_-(K_T) \geq 2$ by exhibiting a downhill subspace; Eq. (S5) is the general exact test.

# SIII. Additional constructions and motif coverage

The balanced-triangle plane uses the first two columns of the basis $\mathbf{p} = (1,-1,0,0)/\sqrt{2}$, $\mathbf{q} = (1,1,-2,0)/\sqrt{6}$, and $\mathbf{r} = (1,1,1,-3)/\sqrt{12}$. With $u, v, w$ as in the main text, its determinant is $[\mathcal{T} - 2\tau(u + v + w) + 3\tau^2]/3$. Therefore Eqs. (7)–(8) are the direct negative-first-entry and positive-determinant test. A sufficiently strong fourth-element interaction may change the remaining curvature but cannot remove the two certified downhill directions.

The full pair-exchange basis contains a third direction, $\mathbf{r} = (0,0,1,-1)/\sqrt{2}$, which exchanges $C$ and $D$. For $P = \Omega_{AB} > 0$ and $J = -\Omega_{CD} > 0$, define $\Xi = (\Omega_{AC} - \Omega_{AD}) - (\Omega_{BC} - \Omega_{BD})$ and $\Phi = (\Omega_{AC} + \Omega_{BC}) - (\Omega_{AD} + \Omega_{BD})$. In the ordered basis $(\mathbf{p}, \mathbf{q}, \mathbf{r})$, the complete 0 K constrained matrix is

$$K_0 = \begin{pmatrix} -P & \Delta/(2\sqrt{2}) & \Xi/2 \\ \Delta/(2\sqrt{2}) & (P - J - \Sigma)/2 & \Phi/(2\sqrt{2}) \\ \Xi/2 & \Phi/(2\sqrt{2}) & J \end{pmatrix}. \tag{S6}$$

Because $-P < 0$ and $J > 0$, this matrix has at least one negative and one positive direction. It has exactly two negative directions if and only if $\det K_0 > 0$. This is an *exact* condition for this stated pair-sign pattern. The main-text plane test is a simpler sufficient condition that fixes the $C - D$ exchange amplitude at zero. At finite equimolar temperature, if $P > \tau$, then $J + \tau > 0$ and $\det(K_0 + \tau I_3) > 0$ is likewise equivalent to exactly two negative modes. This distinction between the sufficient plane and the full three-dimensional condition matters when cross interactions are unequal.

**Two repulsive pairs.** Let $P = \Omega_{AB} > 0$ and $R = \Omega_{CD} > 0$, and exchange the members of each pair using $\mathbf{p} = (1,-1,0,0)/\sqrt{2}$ and $\mathbf{r} = (0,0,1,-1)/\sqrt{2}$. Their coupling is $\Gamma = (\Omega_{AC} - \Omega_{AD} - \Omega_{BC} + \Omega_{BD})/2$. The plane matrix is $\begin{pmatrix} \tau - P & \Gamma \\ \Gamma & \tau - R \end{pmatrix}$, so

$$P > \tau, \qquad R > \tau, \qquad (P - \tau)(R - \tau) > \Gamma^2 \tag{S7}$$

guarantees two downhill directions. Large coupling $|\Gamma|$ can collapse these proposed exchanges into one downhill direction, even though each exchange alone lowers energy.

**Outsider separation.** Choose $\mathbf{p} = (1,-1,0,0)/\sqrt{2}$ and $\mathbf{q} = (-1,-1,3,-1)/\sqrt{12}$. The second vector moves $C$ against the other three elements. The plane entries are

$$\begin{aligned} h_{pp} &= \tau - \Omega_{AB}, \\ h_{qq} &= \tau + [\Omega_{AB} + \Omega_{AD} + \Omega_{BD} - 3(\Omega_{AC} + \Omega_{BC} + \Omega_{CD})]/6, \\ h_{pq} &= [3\Omega_{AC} - \Omega_{AD} - 3\Omega_{BC} + \Omega_{BD}]/\sqrt{24}. \end{aligned} \tag{S8}$$

Applying Eq. (6) to these entries gives another sufficient downhill-plane test. Its composition pattern is distinct from the paired-group plane even if both tests happen to pass for one alloy.

**An entirely attractive chain.** At 0 K, take $\Omega_{AD} = \Omega_{BC} = -b$, $\Omega_{CD} = -c$, with $b, c > 0$, and set the other three interactions to zero for the construction. In the $(\mathbf{p}, \mathbf{q}, \mathbf{r})$ basis used for Eq. (S6), the curvature is

$$K_0 = \begin{pmatrix} 0 & 0 & b \\ 0 & (2b-c)/2 & 0 \\ b & 0 & c \end{pmatrix}. \tag{S9}$$

The $p, r$ block always has one negative direction. The independent $q$ direction is negative when $c > 2b$, so this all-nonpositive interaction pattern has two modes without any repulsive pair. At finite equimolar temperature, the conditions become $c > 2b + 2\tau$ and $b^2 > \tau(c + \tau)$. The strict two-mode state persists under sufficiently small negative perturbations of the zero interactions. The BCC database contains 72 all-negative-interaction quaternaries with two unstable modes at 0 K and none at 500 K.

The motif families overlap. Table S2 gives a mutually exclusive assignment by applying the indicated tests in order. The paired-group row uses the weighted plane test without imposing the illustrative attractive-pair sign $J > 0$. The attractive-chain calculation demonstrates a mechanism but is not a fifth row in this cumulative bookkeeping.

| First satisfied test | New at 0 K | Cumulative at 0 K | New at 500 K | Cumulative at 500 K |
|---|---|---|---|---|
| **Balanced repulsive triangle** | 15,562 | 15,562 | 7,642 | 7,642 |
| **Disjoint pair exchanges** | 2,346 | 17,908 | 983 | 8,625 |
| **Pair exchange and group separation** | 5,403 | 23,311 | 3,338 | 11,963 |
| **Outsider separation** | 2,185 | 25,496 | 937 | 12,900 |
| **Not certified by these four tests** | 1,961 | 27,457 | 955 | 13,855 |

**Table S2. Mutually exclusive assignments among BCC quaternaries with at least two negative modes**. "New" counts alloys not certified by any preceding row, and "cumulative" adds all preceding rows. The four tests certify 92.9% of the 0 K and 93.1% of the 500 K multimode populations. A non-certified alloy may still have two modes; the plane tests are sufficient, not necessary.

Any certified two-dimensional plane can be embedded in an alloy with more than four elements by holding the added fractions fixed. The interaction contribution to its curvature is unchanged. At finite temperature, however, the ideal-entropy curvature depends on the actual fractions of the four changing elements, so the weighted inequalities must be reevaluated at that composition. This is the precise sense in which the construction extends to higher-component alloys.

# SIV. Phase-field models and phase-selected screening

The toy models use three orthonormal conserved composition modes of a quaternary. Their local free-energy density in reduced units is $f = \frac{1}{2}\sum_\alpha \lambda_\alpha \eta_\alpha^2 + \frac{50}{4}(\sum_\alpha \eta_\alpha^2)^2$. The curvature triplets for the three Figure 4 rows are $(-1,1/3,1/3)$, $(-1,-1,3)$, and $(-1,-1/2,2.75)$. Conserved evolution follows $\partial_t \boldsymbol{\eta} = \nabla^2[\partial f/\partial \boldsymbol{\eta} - \kappa \nabla^2 \boldsymbol{\eta}]$ with unit scalar mobility and $\kappa = 4$. Simulations use periodic $128 \times 128$ grids, time step 0.4, and 650 steps, giving reduced time 260. The initial composition noise and observation time are identical across the three rows. The contrast maps and scatter use $x_A - x_B$ and $x_C - x_D$; the scatter contains 3,000 sampled spatial pixels. These simulations establish a visual consequence of the different local curvature spectra within the specified toy dynamics.

The phase-selected screen evaluates the equimolar homogeneous regular-solution free energy at 500 K for each available BCC, FCC, and HCP lattice model. Unary lattice free energies are taken from the supplied unary database, converted to eV per atom, and combined with phase-specific binary interactions and ideal entropy. A system is phase-evaluable if the needed unary energies are available for the compared lattices; 82,251 of 91,390 quaternaries meet the screen's availability criteria. For each system we select the lowest homogeneous lattice energy, then count negative curvatures in that lattice's constrained Hessian. Five exactly-two-mode systems satisfy $D < 0.10$ (Table 1). This comparison selects among homogeneous disordered lattice models; it does not perform a multiphase equilibrium or ordered-compound search.

For Cu–In–Ir–Ti, the complete ideal regular-solution free energy is evolved on a periodic $128 \times 128$ grid with scalar mobility, isotropic $\kappa = 4$, time step 0.1, and 5,000 steps. Zero-sum initial fluctuations have standard deviation 0.0015 and preserve the equimolar mean. Figure 5(b) shows reduced times 290 and 500; only the energy surface in Fig. 5(a) relaxes the stable third mode.